# Trends in Remote Learning-based Google Shopping in the United States due to COVID-19

Isabella Hall[1], Nirmalya Thakur[1], Chia Y. Han[1]

[1] Department of Electrical Engineering and Computer Science, University of Cincinnati, Cincinnati, OH 45221-0030, U.S.A.

## ABSTRACT

The United States of America has been the worst affected country in terms of the number of cases and deaths on account of the severe acute respiratory syndrome coronavirus 2 (SARS-CoV-2) or COVID-19, a highly transmissible and pathogenic coronavirus that started spreading globally in late 2019. On account of the surge of infections, accompanied by hospitalizations and deaths due to COVID-19, and lack of a definitive cure at that point, a national emergency was declared in the United States on March 13, 2020. To prevent the rapid spread of the virus, several states declared stay at home and remote work guidelines shortly after this declaration of an emergency. Such guidelines caused schools, colleges, and universities, both private and public, in all the 50-United States to switch to remote or online forms of teaching for a significant period of time. As a result, Google, the most widely used search engine in the United States, experienced a surge in online shopping of remote learning-based software, systems, applications, and gadgets by both educators and students from all the 50-United States, due to both these groups responding to the associated needs and demands related to switching to remote teaching and learning. This paper aims to investigate, analyze, and interpret these trends of Google Shopping related to remote learning that emerged since March 13, 2020, on account of COVID-19 and the subsequent remote learning adoption in almost all schools, colleges, and universities, from all the 50-United States. The study was performed using Google Trends, which helps to track and study Google Shopping-based online activity emerging from different geolocations. The results and discussions show that the highest interest related to Remote Learning-based Google Shopping was recorded from Oregon, which was followed by Illinois, Florida, Texas, California, and the other states.

# INTRODUCTION

Throughout the last decade (Palvia et al. 2018), but more specifically within the last two years (Moshinski et al. 2021, Moore et al. 2011), society has become far more dependent on remote and online learning technologies for their ability to create flexible, seamless, and interconnected environments for sharing, exchange, communication, and delivery of knowledge and information. Online learning or remote learning, a newer and improved version of distance learning, may broadly be defined as a means of learning that is characterized by the involvement of one or more forms of technology that improves access and availability to educational resources and opportunities for both the learners and the educators (Moore et al. 2011). The recent worldwide emergence of remote learning has been a result of the multitude of advantages that it provides over the traditional approaches of learning. Some of these advantages (James 2021, Zain 2021) include – (1) ease of access, (2) affordability, (3) seamless collaborations from different geographic regions, (4) time saving, and (5) low maintenance cost. The adoption of online learning on a global scale increased at a rate as never seen before over the last couple of years amidst the new normal created by the COVID-19 pandemic (Peimani et al. 2021).

The 2019 Novel Coronavirus (COVID-19), a form of coronavirus that causes respiratory illness of varying degrees in humans, had an outbreak in a seafood and ani-mal market in Wuhan, China, in 2019 (HamiltoncountyhealthOrg 2020). Since then, the virus, characterized by its abilities to mutate and spread rapidly, has affected every country across the world, resulting in infections, critical illness conditions, hospitalizations, and deaths of the order that the world has not seen in a very long time. At the time of writing this pa-per, on a global scale, there have been a total of 247,750,105 infections and 5,018,830 deaths on account of COVID-19 (WorldometersInfo 2021). The World Health Organization has declared this virus as a "pandemic" (WHO 2021), and out of all the countries, the United States has been the worst hit in terms of the number of infections and deaths, with these numbers being 46,864,720 and 766,711, respectively (WorldometersInfo 2021). The COVID-19 virus start-ed spreading in the United States in late 2019, and its spread was characterized by infections, hospitalizations, urgent care treatments, and deaths in magnitudes that the United States economy has never had to deal with in almost about a century (Xiu et al. 2021). As a response to the situation, the United States declared a state of national emergency on March 13, 2020 (WhitehouseGov 2021). The declaration of national emergency was accompanied by the guidelines from the Center for Disease Control (C.D.C.) to work from home to reduce the spread of the virus. Following these guidelines, schools, colleges, and universities, both private and public, across the United States either shut down or switched to remote or distance learning (CDC 2021). In Fall 2019, there were 7,313,623 students enrolled in one or more distance education courses at degree-granting postsecondary institutions in the United States (NCES 2021), and by 2020, 98% of the universities in the United States switched to online mode of instruction (ThinkimpactCom 2021). With the fast transition to remote learning, there was a surge of online shopping for remote learning software, systems, applications, and materials (Kur et al. 2021) that were needed to make the remote learning environments as effective and hands-on as possible for both students and educators.

According to (ThinkimpactCom 2021), since the beginning of the pandemic, 80% of the schools in the United States have purchased or are preparing to purchase additional technology for students. Students across the United States faced multiple issues adapting to completely online forms of education as per this new norm. One of the largest issues was that many students did not have access to the internet or a computer at home. According to (Morgan 2020), around 25% of youth from households that had an annual income of less than $30,000 did not have access to a computer at home. This issue created a larger gap in the equality of information and learning that students received both in the same school as well as across different schools in the same state. The cities and metropolitans simply could not keep up with the demand for remote learning technologies. In a study, it was found that 22% of families incurred financial costs to support remote learning, and the majority of those families were already dealing with lower/lack of income due to COVID-19 (Becker et al. 2020). With the increase of financial costs rising for online learning, it was also found that only 59% of school services that were being received before the pandemic were maintained throughout the online learning process, which increased the need for cheaper and more accessible remote learning methods (Becker et al. 2020). These needs and demands across different states in the United States resulted in more searches on Google than ever before related to the purchase of both new and used technologies, applications, products, and programs related to remote learning. Studying, investigating, and analyzing such web behavior data on Google is expected to provide insight on the trends of adoption of remote learning in different states of the United States due to COVID-19. Such an analysis is also expected to provide insight into the degree of success of switching to online learning during the COVID-19 pandemic across the United States. Investigating this research challenge by exploring the intersections of Big Data, Natural Language Processing, Information Retrieval, Data Science, and Human-Computer Interaction serves as the main motivation for this work.

# LITERATURE REVIEW

The COVID- 19 pandemic has resulted in a seismic shift in the world of online education. The Education Sector of the United States, much like all other countries of the world, is still dealing with this crisis (Ronkowitz et al. 2021). Before COVID-19, two previous epidemics occurred in the 20th and 21st centuries in the United States, during which the education sector was impacted: the 1918 pandemic flu and the 2009 outbreak of H1N1 (swine flu). The average student population in schools and universities was lower during these two pandemics (Crocker 2021). Classes could be properly restructured and expanded to include distant learning even if students choose to isolate themselves on campus. The morale of the people also played a role in 1918 during the outbreak of the flu (Cozens 2020).

An overwhelming majority of the population was engaged in providing various forms of service to World War I at the time. The sacrifices of a majority of the population were greater than the sacrifices made by others whose education was impeded. Recently, a number of studies have attempted to determine the most important considerations and best practices that contribute to the acceptability, assimilation, and success of online education (Asgari 2021) as the switch to online education has been on a wide scale in almost all countries and geographic regions of the world on account of COVID-19. These factors included course design, course content support, instructor's characteristics, and students' familiarity with and access to technical resources. Faculty and students at academic institutions in different geographic regions of the world, specifically the United States, that were primarily focused on conventional face-to-face education faced several problems as a result of the rapid transition to online training, which was prompted by the rapid surge of COVID-19 (Almaiah et al. 2019, Al-Gahtani et al. 2016, Almaiah 2020). There was a growing amount of research on the educational consequences of COVID-19 as the outbreak progressed in different states. In June 2020, a team of researchers (Saw et al. 2020, Saw et al. 2020) performed a countrywide survey of STEM professors and students in the United States to evaluate the impact of COVID-19 on the education system. According to their research, 35.5% of doctorate students, 18% of master's students, and 7.6% of undergraduate students missed their graduation due to COVID-19. In addition to challenges centered around graduation, students and professors in the United States experienced a multitude of diverse problems and worries as they considered whether and how to continue or not continue online education in light of the surge in COVID-19 cases, as reported in several blogs posts, editorials, and short reports (EducauseEdu 2021, HomonymCa 2020, InsidehigheredCom 2021) during the initial months of the outbreak. Topics included how to deal with inequalities, needs, how to teach online, how to share faculty and student experiences, and the long- and short-term consequences for institutions (Forbes Magazine 2020, Harv Bus Rev 2020). These blog posts, editorials, and short reports have prompted researchers in the last few months to perform comprehensive studies on these problems and the associated needs faced by the students, professors, and institutions in different regions of the United States (Nguyen et al. 2020, Quezada et al. 2020. Iyer et al. 2020. Mercier et al. 2021. Basilaia et al. 2021).

All these works indicate a similar trend of needs centered around familiarization with online learning or re-mote learning-based tools, applications, and products expressed by students, professors, and institutions from different states and regions of the United States. However, there has not been any prior work conducted in this field thus far that attempts to investigate, analyze, interpret, and quantify such needs and the trends in the same across different regions in the United States on account of COVID-19. Therefore, this work aims to address this knowledge gap in this field of research.

## PROPOSED APPROACH

We performed this study using Google Trends (Google 2020). Google Trends is a platform by Google that allows the web mining of search interests related to a specific set of keywords by performing semantic analysis of relevant search queries on Google. These results can be mined at a global scale as well as they can be filtered down at a country, state, and city level. Google Trends also provides multiple ways to filter the search interest data, which includes location, time period, search category type, and type of search - such as an image search, news search, etc. With Google Trends, one can also compare search results from multiple searches to look at common occurrences, and differences within the information pulled from the associated Google searches. This helps further analyze the data to better comprehend the larger picture and what the actual data fully represents in the grand scheme of the project. Google Trends uses a scale from 0 to 100 to represent this popularity in terms of search interests. This numerical value represents a weighted result based on the topic's pro-portion to all searches on all topics recorded on Google (Google Trends 2020). The number from 0 to 100 is determined by dividing the data point by the total searches of the area and the time period that it represents so that the algorithm is able to compare the relative popularity of the search terms and assign it an appropriate number in this scale. Figure 1 shows a screenshot from the Google Trends website with the implementation of the proposed approach. As shown in Figure 1, we entered the keywords as "remote learning", selected "Google Shopping" to filter out all those search results on Google which dealt with shopping activities, the region as "United States", and duration as "03/13/2020 – 10/21/2021". Here, March 13, 2020, was selected as the start date for this study as the national emergency on account of COVID-19 was declared on this day in the United States. The end date for this study was selected as the date on which this study was performed to accommodate for the maximum amount of Google Shopping-based Big Data that was available on Google since the declaration of the emergency.

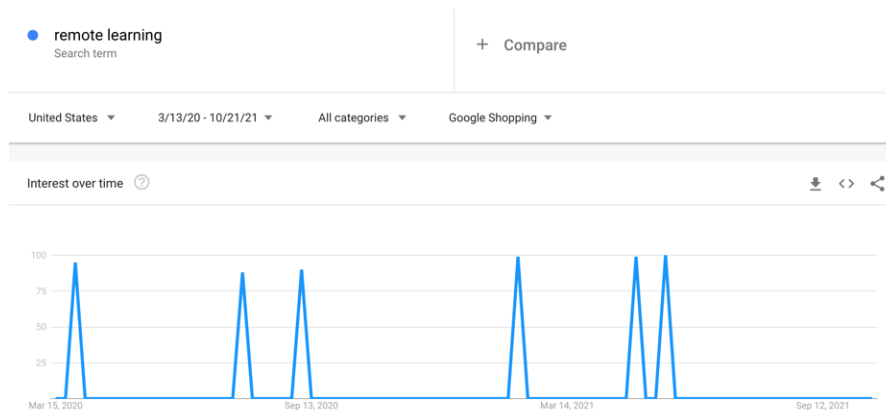

Figure 1. Screenshot from Google Trends that shows the implementation of this approach along with the keywords and the other specifications of the study.

## RESULTS AND DISCUSSIONS

The above methodology provided our results in terms of search interest values in the context of Google Shopping-based web behavior data on a scale of 0-100 for the time period of March 13, 2020, to October 21, 2021, for all the states across the United States. For each state, the search interest value was normalized to fall within the 0-100 scale by Google Trends, as explained in Section 3. States from which a significant amount of relevant search interests were not recorded were assigned a 0 value on this scale. Figure 2 shows the ranking of the states from which significant relevant search interest values were recorded in the above-mentioned time period. As can be seen from Figure 2, Oregon recorded the highest search interest value, which was followed by Illinois, Florida, Texas, California, and the other states. These trends reflect the Google Shopping-based behavior related to the purchase of online learning-based technologies, applications, products, and programs originating from each of these states. The numerical values represent the magnitude of user interest which helps us to potentially conclude that the maximum Google Shopping-based behavior related to the purchase of online learning-based technologies, applications, products, and programs has been observed in Oregon so far since the out-break of COVID-19 in the United States. In addition to directly reflecting the potential sales, these results can also be interpreted to study the associated market potential, sales outlook, and consumer perspectives of online learning-based technologies, applications, products, and programs in different states across the United States.

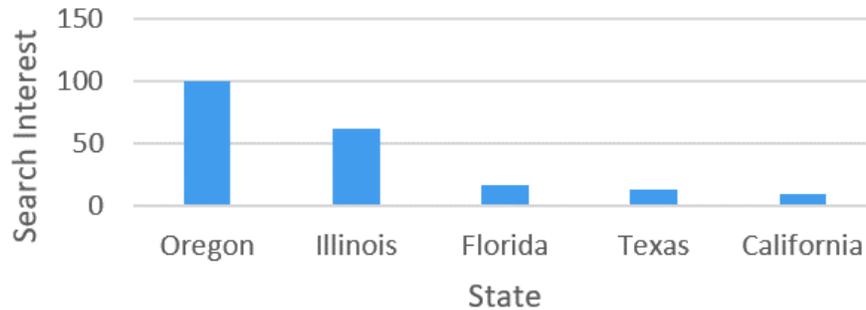

Figure 2. Analysis of the search interest values related to Google Shopping originating from different states (only those states that recorded significant values) of the United States.

## CONCLUSION AND FUTURE WORK

The declaration of national emergency on March 13, 2020, in the United States to reduce the spread of COVID-19 was associated with work from home guidelines which cased private and public schools, colleges, and universities in all the states in the United States to switch to remote or online teaching methods for an extended period of time. On account of the same, Google experienced an increase in online shopping for remote learning-based technologies, applications, products, and programs by both educators and students, as both these user groups strived to address their needs and demands associated with adjusting to this new norm of education. This resulted in the generation of Big Data from the relevant web behavior on Google since the outbreak of this pandemic in the United States. In this paper, we have presented an approach to investigate, analyze, quantify, and interpret these trends of Google Shopping related to remote learning that emerged since March 13, 2020, on account of COVID-19 from all over the 50-United States. This interdisciplinary study at the intersections of Big Data, Natural Language Processing, Information Retrieval, Data Science, and Human-Computer Interaction, was performed using Google Trends. The results presented and discussed uphold the relevance of this study towards the interpretation of Google Shopping-based web behavior related to the purchase of remote learning-based technologies, applications, products, and programs originating from each of these states. The results also discuss the associated trends in Google Shopping-based web behavior by comparing the magnitude of the associated search interests. Finally, the interpretation of the results also indicates the potential of this approach for studying market potential, sales outlook, and consumer perspectives of remote learning-based technologies in different geographic regions of the United States. Future work in this field would involve performing the study in other countries such as India, Brazil, the United Kingdom, and Russia, which are amongst the worst-hit countries by COVID-19.

# AUTHOR CONTRIBUTIONS

Conceptualization, N.T. and I.H.; Methodology, I.H., and N.T.; Data Curation, I.H.; Formal Analysis, I.H.; Data Visualization and Interpretation, I.H.; Results, I.H.; Discussion of Results, N.T.; Writing-Original Draft Preparation, I.H., and N.T.; Writing-Review and Editing, N.T; supervision, N.T; project administration, N.T and C.Y.H.; Funding Acquisition, Not Applicable. All authors have read and agreed to the published version of the manuscript.